\documentclass[journal,12pt,onecolumn,draftclsnofoot]{IEEEtran}

\usepackage{color}
\usepackage{multirow}
\usepackage{graphicx}
\usepackage{epstopdf}
\usepackage[cmex10]{amsmath}
\usepackage{amssymb}

\usepackage{multirow}
\usepackage{amsthm}
\usepackage{cite}
\usepackage{acronym} %This package will write out the acronym the first time it is used and write it as an acronym for the rest of the paper
\usepackage{algorithm, algcompatible}

\algnewcommand\INPUT{\item[\textbf{Input:}]}
\algnewcommand\OUTPUT{\item[\textbf{Output:}]}

\allowdisplaybreaks

\begin{document}

\title{Extreme Age of Information for Wireless-Powered Communication Systems} 
\author{Nikolaos~I.~Miridakis,~\IEEEmembership{Senior Member,~IEEE}, Zheng~Shi,~\IEEEmembership{Member,~IEEE},\\ Theodoros~A.~Tsiftsis,~\IEEEmembership{Senior Member,~IEEE}, and Guanghua~Yang,~\IEEEmembership{Senior Member,~IEEE}
%\thanks{Copyright \copyright  2021 IEEE. Personal use of this material is permitted. However, permission to use this material for any other purposes must be obtained from the IEEE by sending a request to pubs-permissions@ieee.org.}
\thanks{\textit{Corresponding Author: T. A. Tsiftsis.}}
\thanks{The authors are with the School of Intelligent Systems Science \& Engineering, Jinan University, Zhuhai Campus, Zhuhai 519070, China. N. I. Miridakis is also with the Dept. of Informatics and Computer Engineering, University of West Attica, Aegaleo 12243, Greece (e-mails: nikozm@uniwa.gr, zhengshi@jnu.edu.cn, theo\_tsiftsis@jnu.edu.cn, ghyang@jnu.edu.cn).}
}

%\markboth{}{}

\maketitle

\begin{abstract}
The extreme or maximum age of information (AoI) is analytically studied for wireless communication systems. In particular, a wireless powered single-antenna source node and a receiver (connected to the power grid) equipped with multiple antennas are considered when operated under independent Rayleigh-faded channels. Via the extreme value theory and its corresponding statistical features, we demonstrate that the extreme AoI converges to the Gumbel distribution whereas its corresponding parameters are obtained in straightforward closed-form expressions. Capitalizing on this result, the risk of the extreme AoI realization is analytically evaluated according to some relevant performance metrics, while some useful engineering insights are manifested.  
\end{abstract}


\IEEEpeerreviewmaketitle

\section{Introduction}
\IEEEPARstart{A}{ge} of information (AoI) represents an insightful performance metric for wireless communication systems, which indicates the timeliness of data delivery. Doing so, it has numerous applications in a set of various recent and upcoming real-time networking paradigms, e.g., Internet-of-things, vehicular and machine-type communications, healthcare, infrastructure and environment monitoring, as well as low-latency and delay-critical services \cite{j:RoyYates2021,j:Yates2020}. Source nodes in these applications are usually simple sensor devices which are battery-powered and thus become energy-constrained. The energy efficiency can be enhanced, when energy harvesting is implemented at the considered system nodes. In particular, energy harvesting via wireless power transfer (WPT) can be used for enabling perpetual operation of wireless devices. In fact, WPT has attracted increasing interests mainly due to its capability of converting received RF signals into electricity, which in turn is able to provide stable and controllable power to prolong the lifetime of low-power energy-constrained autonomous networks \cite{j:Kriki2019}.

In WPT systems, AoI plays a pivotal role since the source nodes are in principle power-limited and thus the scenario of erroneous detection (which is accompanied with consecutive retransmissions) dramatically affects the freshness of information at the receiver. To this end, various research studies have analyzed the impact of AoI in such systems; see, e.g., \cite{j:Kriki2019,c:Sleem2020,j:JiaCao2021} and relevant references therein. Nonetheless, most studies so far have focused on the analysis and/or optimization of the average AoI (or peak AoI). Although the average AoI defines a key statistical performance metric, it can not capture the effect of information aging for \emph{extremely} rare events. A paradigm of particular interest is the maximum AoI during the communication process of the transceiver, which is located in the right tail of the (instantaneous) AoI distribution. The extreme (or maximum) AoI may influence more drastically the system performance rather than its average counterpart in certain applications; such as autonomous vehicles, tactile Internet as well as other exciting yet delay-critical and low-latency services. To our knowledge, the effect of extreme AoI has only been analytically studied in \cite{j:LiuFeng2019} and \cite{j:Aziz2020}. In the latter works, however, the (conditional on some time threshold) extreme AoI was approached by the generalized extreme value distribution. Yet, its corresponding parametric values were numerically computed, while the energy limitation of the supporting source nodes was not considered therein.

Capitalizing on the aforementioned observations, for the first time in this Letter, the (unconditional) extreme AoI is analytically studied for WPT systems. Using the statistical tools of the extreme value theory, we explicitly indicate that the extreme AoI is suitably approached by the Gumbel distribution. The parametric values of the said distribution are derived in simple closed-form expressions. Furthermore, the \emph{risk} arising from an extremely high aging of information is analytically quantified via some relevant performance metrics; namely, the value-at-risk (VaR) and conditional VaR (CVaR). With the latter metrics at hand, useful engineering insights are revealed which may become impactful for the system design of delay-critical and/or low-latency wireless networking applications.  

{\it Notation:} $\mathbb{E}[\cdot]$ is the expectation operator; ${\rm Pr[\cdot]}$ stands for the probability operator; $|\cdot|$ represents the absolute (scalar) value operator; $y|z$ denotes that $y$ is conditioned on $z$ event; $\hat{x}$ denotes an estimate of $x$; ${\rm li}(\cdot)$ is the logarithmic integral function \cite[Eq. (4.211.2)]{tables}; $\Gamma(\cdot,\cdot)$ is the upper incomplete Gamma function \cite[Eq. (8.350.2)]{tables}; $\epsilon$ denotes the Euler-Mascheroni constant \cite[Eq. (8.367)]{tables}.

\section{System Model}
\begin{figure}[!t]
\centering
\includegraphics[trim=0.5cm 0.0cm 0.5cm .1cm, clip=true,totalheight=0.20\textheight]{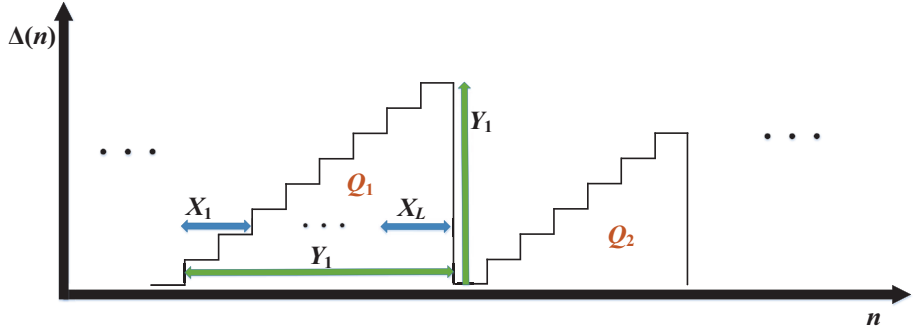}
\caption{The AoI progression of the considered system model.}
\label{fig1}
\end{figure}

Consider a single-antenna source node that generates status updates and sends them at a given destination equipped with $N\geq 1$ antennas {\color{black}which is connected to the power grid}. The transmitting source has energy harvesting capabilities {\color{black}and fills its battery via WPT from a dedicated energy transmitter (ET)}; it stores energy in a capacitor of size $S$ and transmits fresh data only when its battery is full. Then, all its harvested energy is used striving for a successful data reception. It is assumed that ET is connected to the power grid, whereas it continuously broadcasts a certain RF signal with a fixed power $P_{\rm t}$. {\color{black}Thereupon, assuming a linear WPT model, the stored energy of the source at the $l^{\rm th}$ timeslot is modeled by $E_{l}\triangleq \min\{E_{l-1}+\eta P_{\rm t}|u_{l}|^{2},S\}$ when $E_{l-1}<S$; or $E_{l}\triangleq \min\{\eta P_{\rm t}|u_{l}|^{2},S\}$ when $E_{l-1}\geq S$. Also, $\eta$ represents the RF-to-DC conversion efficiency, while $u_{l}$ denotes the channel fading coefficient for the link between the source node and ET at the $l^{\rm th}$ timeslot. Although a sophisticated nonlinear WPT modeling is more appropriate for realistic conditions, the adopted simplified linear model serves as a lower performance bound on the nonlinear harvested energy used in practice \cite{j:Kriki2019,j:MiridakisTsif2018}}. In addition, the energy and communication links are assumed orthogonal (e.g., occupy different frequency bands) so as to avoid mutual interference. Further, the data communication is achieved in consecutive timeslots with a slot size of one time unit.\footnote{Henceforth, the energy and power metrics become equal and will be used interchangeably.} In Fig.~\ref{fig1}, the AoI evolution across time is illustrated for the considered WPT system model. 

All the included wireless links experience independent quasi-static Rayleigh fading conditions, whereon the channel remains fixed for the duration of one timeslot while it changes afterwards. At the receiver, with a perfect channel-state information (CSI) at hand, maximum ratio combining (MRC) is utilized in order to enhance the received channel gains. Consequently, the received signal-to-noise ratio (SNR) at the $i^{\rm th}$ timeslot reads as
\begin{align}
\gamma_{i}=\frac{S}{N_{0}}\sum^{N}_{v=1}|h_{i,v}|^{2},
\label{snr}
\end{align}
where $N_{0}$ is the additive white Gaussian noise power and {\color{black}$h_{i,v}$ denotes a zero-mean and unit-variance circularly symmetric complex Gaussian fading coefficient at the $i^{\rm th}$ timeslot and $v^{\rm th}$ received antenna element}. In Fig.~\ref{fig1}, $X_{i}$ represents the number of consecutive timeslots required for the capacitor being fully charged. When the energy level reaches the $S$ value, a signal transmission occurs; in the case when a successful reception occurs, ${\rm log}_{2}(1+\gamma_{i})\geq R$ is satisfied with $R$ standing for a predetermined data rate (in bps/Hz) and the AoI is reset to zero. Otherwise, the capacitor starts to recharge in the following timeslots and then it repeats the signal retransmission until a successful transmission occurs. In the latter case, all the timeslots required for a successful reception (i.e., the interarrival time between consecutive decoded messages) are concretely defined as $Y_{j}\triangleq \sum^{L}_{i=1}X_{i}$ with $L<\infty$ (as per Fig.~\ref{fig1}).

\section{Performance Metrics}
The instantaneous AoI at a given timeslot $m$ is defined as 
\begin{align}
\Delta(m)\triangleq m-\mathcal{G}(m), 
\label{instAoI}
\end{align}
where $\mathcal{G}(m)$ is the generation time of the last update message.

\subsection{Average AoI}
For a \emph{stationary and ergodic} information updating process, the time average AoI tends to its ensemble average counterpart given an asymptotically high number of consecutive samples. Therefore, it holds that 
\begin{align}
\overline{\Delta}&\triangleq \frac{1}{M}\sum^{M}_{m=1}\Delta(m)=\frac{K}{M}\frac{1}{K}\sum^{K}_{i=1}Q_{i}=\frac{\mathbb{E}[Q]}{\mathbb{E}[Y]},\quad M\rightarrow +\infty,
\label{avAoI}
\end{align}
{\color{black}where $\mathbb{E}[Q]\triangleq \frac{1}{K}\sum^{K}_{i=1}Q_{i}$ and $1/\mathbb{E}[Y]$ denotes the steady-state rate of the packet arrival time, such that $\mathbb{E}[Y]\triangleq \lim_{M\rightarrow +\infty}\frac{M}{K}=\frac{1}{K}\sum^{K}_{i=1}Y_{i}$}. Also, $Q_{i}$ denotes the disjoint area in Fig.~\ref{fig1} which is computed as
\begin{align}
Q_{i}=\sum^{Y_{i}}_{u=1}u=\frac{1}{2}\left[Y_{i}(Y_{i}+1)\right].
\label{Q}
\end{align}
It turns out that \cite[Thm.~1]{j:Kriki2019} 
\begin{align}
\overline{\Delta}=\frac{1}{2}\left(\frac{\mathbb{E}[Y^{2}]}{\mathbb{E}[Y]}+1\right)=\frac{1+3g+g^{2}}{2 (g+1)}+\frac{(1+g) p}{1-p}+\frac{1}{2},
\label{avAoIcf}
\end{align}
where 
\begin{align}
p\triangleq{\rm Pr}[{\rm log}_{2}(1+\gamma_{i})<R]=1-\frac{\Gamma\left(N,\frac{w d^{\alpha}_{c}(2^{R}-1)}{S/N_{0}}\right)}{(N-1)!}
\label{out}
\end{align}
is the outage probability with $w$, $d_{c}$ and $\alpha$ denoting the propagation attenuation at a reference distance of 1 m., transceiver distance of the communication link and path-loss exponent, respectively. Also, 
\begin{align}
g\triangleq \frac{w d^{\alpha}_{e} S}{\eta P_{\rm t}}, 
\end{align}
where $d_{e}$ is the transceiver distance of the WPT link. 

For the limiting case when $N\rightarrow +\infty$ (i.e., a massive receive antenna array) and with the aid of \cite[Eq. (8.11.5)]{b:NIST}, $p\rightarrow 0^{+}$ and \eqref{avAoIcf} simplifies to $\overline{\Delta}^{(N\rightarrow +\infty)}\approx (2+4g+g^{2})/(2+2 g)$. Moreover, in the asymptotically high transmission power regime (where $S/P_{\rm t}\rightarrow 0^{+}$ and thus $g\rightarrow 0^{+}$), the average AoI tends to one timeslot.\footnote{To this end, the simultaneous energy harvesting and data transmission is required. Doing so, the \emph{energy half-duplex problem} can be avoided by assuming two antennas and two energy storage devices (one used for data transmission and the other for energy harvesting, correspondingly), as in \cite{j:Kriki2019} and \cite{j:MiridakisTsif2018}.}

\subsection{AoI Variance}
Following similar lines of reasoning as for the derivation of \eqref{avAoI}, we get the AoI variance derived by
\begin{align}
\nonumber
&\mathcal{V}_{\Delta}=\frac{\mathbb{E}[Q^{2}]}{\mathbb{E}[Y]}-\overline{\Delta}^{2}=\frac{\mathbb{E}\left[\left[Y(Y+1)\right]^{2}\right]}{4 \mathbb{E}[Y]}-\frac{\mathbb{E}^{2}\left[Y(Y+1)\right]}{4 \mathbb{E}^{2}[Y]}\\
&=\frac{1}{\mathbb{E}[Y]}\left(\frac{\mathbb{E}[Y^{4}]}{4}+\frac{\mathbb{E}[Y^{3}]}{2}+\frac{\mathbb{E}[Y^{2}]}{4}\right)-\frac{\left(\mathbb{E}[Y^{2}]+\mathbb{E}[Y]\right)^{2}}{4 \mathbb{E}^{2}[Y]}.
\label{varianceAoI}
\end{align}
The above formula yields a closed-form expression given by
\begin{align}
\nonumber
\mathcal{V}_{\Delta}=&\frac{1}{4 (g+1)^2 (1-p)^3}\bigg[g^5 \left(p^3+11 p^2+11 p+1\right)\\
\nonumber
&+g^4 \left(32 p^2+78 p-2 p^3+12\right)\\
\nonumber
&+g^3 \left(26 p^2+174 p-2 p^3+42\right)\\
\nonumber
&+2 g^2 \left(p^3+5 p^2+89 p+25\right)\\
&+4 g \left(3 p^2+22 p+5\right)+4 p (p+5)\bigg].
\label{varianceAoIcf}
\end{align}
The proof of \eqref{varianceAoIcf} is relegated in the Appendix. For the limiting case when $N\rightarrow +\infty$, $p\rightarrow 0^{+}$ and \eqref{varianceAoIcf} simplifies to
\begin{align}
\mathcal{V}^{(N\rightarrow +\infty)}_{\Delta}\approx \frac{g \left(g^4+12 g^3+42 g^2+50 g+20\right)}{4 (g+1)^2}.
\label{varianceAoIapprox}
\end{align}
Further, in the asymptotically high transmission power regime (where $g\rightarrow 0^{+}$), the AoI variance tends to zero, as expected. 

\subsection{Extreme AoI Statistics}
Although the AoI variance is a powerful statistical tool and (in contrast to the average AoI) provides useful information on the fluctuation of AoI, it is not sufficient to estimate the behavior of extreme cases (e.g., the maximum attainable AoI). To this end, we resort to the extreme value theory which efficiently describes rare events. Subsequently, we show that the Gumbel distribution suitably models the statistical behavior of extreme (rare) AoI events for the considered system model.

We commence by recognizing that $X_{i}$ (i.e., the number of consecutive timeslots required for a fully-charged capacitor) follows a Poisson distribution as per \eqref{pmfX}. Then, the $j^{\rm th}$ interarrival time between two successive correctly decoded update messages (say, the instantaneous AoI) $Y_{j}=\sum^{L}_{i} X_{i}$ with $L< \infty$ is also a Poisson RV with a corresponding PMF (conditioned on $L$ attempts) being
\begin{align}
{\rm Pr}[Y_{j}=m|L]=\exp(-L g)\frac{(L g)^{m-1}}{(m-1)!}.
\label{pmfY}
\end{align}
According to the extreme value theory \cite{b:Haan2006}, the maximum $Y_{j}$ defined as $Y_{\max}\triangleq \max \{Y_{j}\}^{\infty}_{j=1}$ belongs to the maximum domain of attraction (MDA) of one of the three possible distribution types; namely the Gumbel, Fr\'echet or Weibull. According to \cite[Thm.~3]{j:SHIMURA2012}, the Poisson distribution is recoverable\footnote{The explicit definition of a recoverable distribution is provided in \cite[Eq. (1.1)]{j:SHIMURA2012}.} in the MDA of Gumbel distribution yet not uniquely recovered. This occurs due to the discrete nature of Poisson samples. Nevertheless, for a sufficiently high parameter, say $L g>10$ in the parent PMF of $Y_{j}$ as per \eqref{pmfY}, Poisson distribution resembles the Gaussian distribution. As shown in the numerical results section, the assumption of $L g>10$ is indeed a reasonable assumption since it reflects parametric values of practical interest. Insightfully, the latter distribution belongs to the MDA of Gumbel distribution and is uniquely recovered therein \cite[Thm.~5]{j:SHIMURA2012}. Hence, $Y_{\max}$ converges to the Gumbel distribution.

The Gumbel distribution function is defined as
\begin{align}
F_{\rm G}(x;\mu_{\rm G},\sigma_{\rm G})=\exp\left(-\exp\left(-\frac{x-\mu_{\rm G}}{\sigma_{\rm G}}\right)\right),
\label{cdfGumbel}
\end{align}
where $\mu_{\rm G}$ and $\sigma_{\rm G}$ are the so-called location and scale parameter, respectively. Also, the expected value and variance of Gumbel distribution are given as $\mu_{\rm G}+\sigma_{\rm G} \epsilon$ and $\sigma^{2}_{\rm G} \pi^{2}/6$, correspondingly. Via the method of moments-matching estimation, we obtain their respective estimates as
\begin{align}
\hat{\mu}_{\rm G}=\overline{\Delta}-\epsilon \frac{\sqrt{6 \mathcal{V}_{\Delta}}}{\pi},
\label{mG}
\end{align}
and 
\begin{align}
\hat{\sigma}_{\rm G}=\frac{\sqrt{6 \mathcal{V}_{\Delta}}}{\pi}.
\label{sG}
\end{align}
Thus, the distribution of the extreme (i.e., maximum) AoI, $\Delta_{\max}$, is approached by
\begin{align}
F_{\Delta_{\max}}(x)\approx F_{\rm G}(x;\hat{\mu}_{\rm G},\hat{\sigma}_{\rm G}).
\label{cdfMax}
\end{align}

Provided with the tail statistics of AoI, $\Delta_{\max}$ can be further quantified given a specific \emph{level of risk}. By introducing the notion of confidence level $a$, we are able to characterize the level of risk in terms of AoI; for instance, a confidence level of $a$ means that there is an $a\%$ possibility that the worst-case AoI will not exceed a certain time threshold. Typical percentile values of the confidence level are $a=99\%$ or $a=95\%$. VaR is regarded as a key performance metric to quantify the effect of rare events. It is defined as \cite[Def.~3.3]{j:Artzner1999}
\begin{align}
\nonumber
{\rm VaR}_{\Delta_{\max}}(a)&\triangleq Q_{\Delta_{\max}}(a)\\
&\approx \hat{\mu}_{\rm G}-\hat{\sigma}_{\rm G}{\rm ln}\left(-{\rm ln}(a)\right),\quad 0<a<1,
\label{VAR}
\end{align}
where $Q_{\Delta_{\max}}(\cdot)$ is the quantile function of $\Delta_{\max}$. However, VaR is an incoherent risk measure.\footnote{A coherent risk metric satisfies certain properties regarding its supporting function; namely, translational invariance, sub-additivity, monotonicity, and homogeneity \cite{j:Artzner1999}.} An alternative measure, which is indeed coherent and more reliable than VaR is the so-called CVaR. It is presented as \cite[Prop.~15]{j:Norton2021}
\begin{align}
\nonumber
{\rm CVaR}_{\Delta_{\max}}(a)&\triangleq \frac{1}{1-a}\int^{1}_{a}{\rm VaR}_{\Delta_{\max}}(a)dy\\
&\approx \hat{\mu}_{\rm G}+\frac{\hat{\sigma}_{\rm G}}{1-a}\big[{\rm li}(a)-a {\rm ln}\left(-{\rm ln}(a)\right)+\epsilon \big].
\label{CVAR}
\end{align}
It is noteworthy that ${\rm CVaR}_{\Delta_{\max}}(a)$ is the expectation on the worst $(1-a)\%$ values of $\Delta_{\max}$, which reflects that it is more sensitive than VaR to the shape of the AoI distribution in the right tail. Notably, the term in brackets within \eqref{CVAR} can be \emph{a priori} computed and therefore is regarded as an offline operation.

\section{Numerical Results and Discussion}
The derived analytical results are verified via numerical validation, whereas they are cross-compared with corresponding Monte-Carlo simulations. Without loss of generality and for ease of presentation, $d_{e}=d_{c}\triangleq d$ is considered. Also, the RF-to-DC conversion efficiency is set to be $\eta=0.5$ while $w=10^{3}$.

\begin{figure}[!t]
\centering
\includegraphics[trim=3.0cm .5cm 3.0cm .2cm, clip=true,totalheight=0.3\textheight]{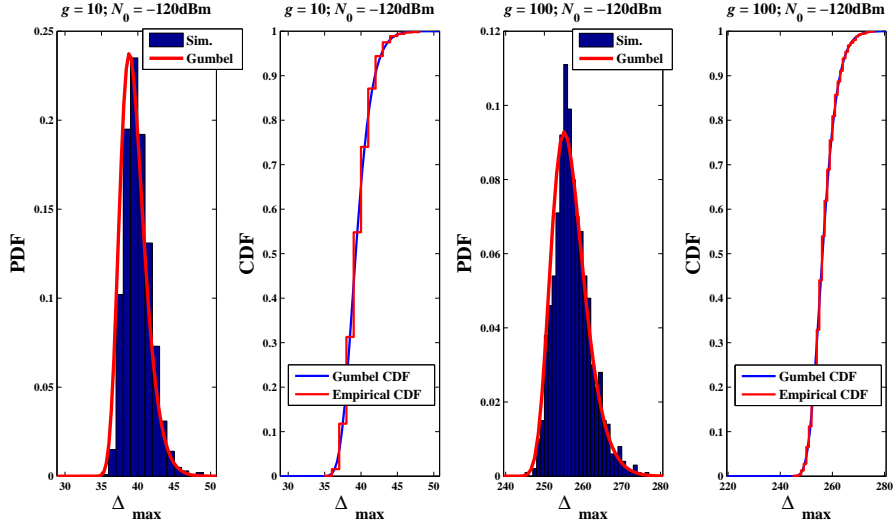}
\caption{Simulation data (entitled as `{\rm Sim.}' in legend) of the maximum AoI vs. the considered Gumbel distribution as per \eqref{cdfMax} for different system settings and a single-antenna transceiver with a target data rate $R=0.1$ bps/Hz. PDF and CDF denote the probability density function and cumulative distribution function, respectively.}
\label{fig2}
\end{figure}

In Fig.~\ref{fig2}, $g=10$ and $g=100$ correspond to a capacitor size $S=0.2\times 10^{-3}$, path-loss exponent $\alpha=2.4$, transmit power $P_{\rm t}=40$dBm as well as $d=10$m. and $d=26$m., respectively. Simulation data are cross-compared with the considered Gumbel distribution. Obviously, there is an almost perfect fit at extremely rare events in the right tail when maximum AoI is realized.

\begin{figure}[!t]
\centering
\includegraphics[trim=2.0cm .2cm 3.0cm .2cm, clip=true,totalheight=0.3\textheight]{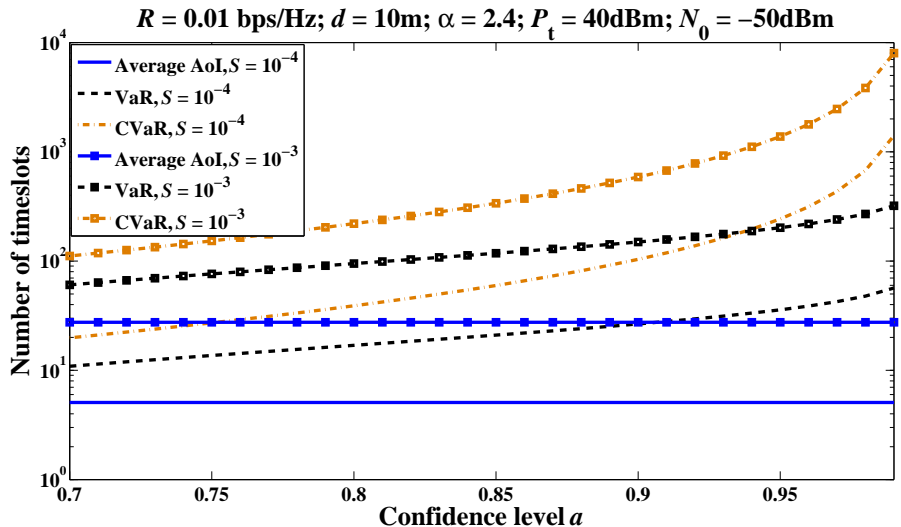}
\caption{Performance comparison of the average AoI, VaR and CVaR vs. various confidence levels for different system settings and a single-antenna transceiver.}
\label{fig3}
\end{figure}

In Fig.~\ref{fig3}, the performance of average AoI, VaR and CVaR is illustrated against different confidence levels. For all the considered cases, CVaR is higher than VaR since the former metric can capture the scenario of extreme AoI more rigorously than the latter one. Both the said metrics maintain a notable difference from the average AoI (with respect to the number of required timeslots), even in the low confidence percentile of $70\%$. In fact, the latter difference becomes even more pronounced as the confidence level increases; this implies that for an almost sure convergence (i.e., for $a=99\%$), taming quite a low AoI becomes a challenging task. Even for a more relaxed confidence level (e.g., when $a=85\%$ or $a=90\%$), there is a remarkable deviation between the most likely AoI (say, the average AoI) and its worst case counterpart. Moreover, the case when the capacitor size is $S=10^{-4}$ reflects on a superior performance in terms of AoI in contrast to the case when $S=10^{-3}$. It seems that (for this particular system setting) is better to operate using a lower transmission power profile at the source node so as to achieve a faster capacitor charge, which in turn reduces the total AoI. However, this holds true due to the moderately high received channel gains (i.e., a short link distance, $d=10$m.) implying that the outage probability $p$ is rather low.

\begin{figure}[!t]
\centering
\includegraphics[trim=2.0cm .1cm 2.5cm .2cm, clip=true,totalheight=0.3\textheight]{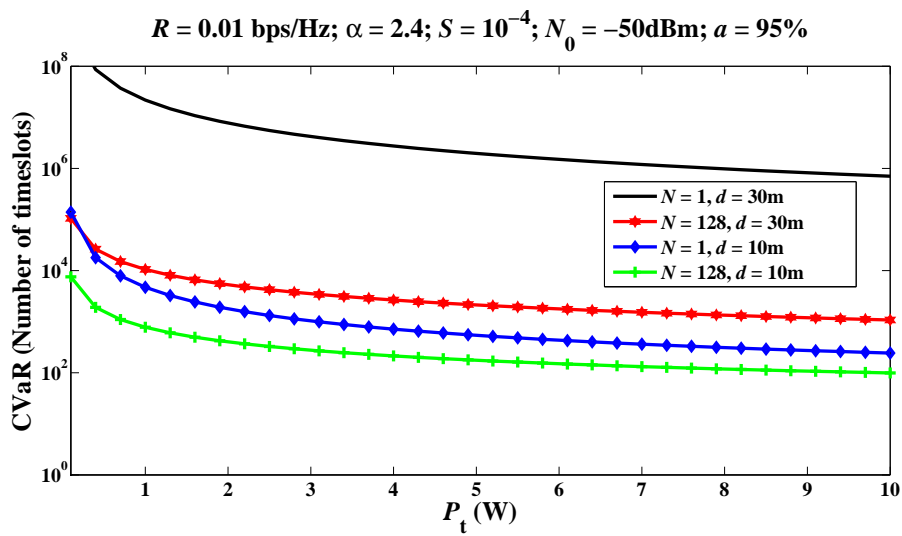}
\caption{Performance of CVaR vs. various transmission power levels (of the energy transmitter) for different system settings.}
\label{fig4}
\end{figure}

Evidently, a promising solution to the aforementioned challenge (whenever the received channel gains are severely reduced) is to provide multiple antennas at the receiver side, as illustrated in Fig.~\ref{fig4}. In order to obtain such an insight more emphatically, a massive antenna scale is applied. As expected, CVaR is reduced for the multiple antenna case since the received channel gains are being enhanced for higher $N$ values, which in turn reflects on less packet errors and thus a shorter AoI. As demonstrated in Fig.~\ref{fig4}, the performance difference is more impactful when channel fading and signal propagation attenuation is more severe (e.g., far-distant communication links). 

\section{Conclusions}
The extreme AoI of a WPT communication system was analytically studied under independent Rayleigh channel fading conditions. It turns out that the extreme AoI of the considered system is suitably approached by the Gumbel distribution, while its associated parameters depend on the expected value and variance of AoI which were derived in simple closed-form expressions. Moreover, the VaR and CVaR metrics were obtained which demonstrate the deviation of the extreme AoI from its corresponding average counterpart. For a system setting of practical interest, it is observed that the capacitor size directly impacts on the extreme AoI, while an increasing number of receive antennas is beneficial only for severely degraded received channel gains (e.g., far-distant channel links).

\appendix
\label{appa}
\numberwithin{equation}{section}
\setcounter{equation}{0}
According to \eqref{varianceAoI}, the moments-function of $Y$ should be computed. Recall that $Y=\sum_{i} X_{i}$ and $X_{i}$ is a mutually independent RV $\forall i$. The probability mass function (PMF) of $X_{i}$ defines the number of consecutive timeslots required for a fully-charged capacitor and is expressed as \cite[Eq. (14)]{j:Kriki2019}\footnote{Since $X_{i}$ RVs are statistically independent, the $i^{\rm th}$ index is dropped hereinafter.}
\begin{align}
{\rm Pr}[X=k]=\exp(-g)\frac{g^{k-1}}{(k-1)!}.
\label{pmfX}
\end{align}
Hence, the moment-function of $X$ is given by
\begin{align}
\mathbb{E}[X^{n}]\triangleq \sum^{\infty}_{k=1}k^{n}{\rm Pr}[X=k],
\label{momX}
\end{align}
and it is specialized for the first four moments, respectively, as follows
\begin{align}
\nonumber
\mathbb{E}[X]&=1+g,\\
\nonumber
\mathbb{E}[X^{2}]&=1+3 g+g^{2},\\
\nonumber
\mathbb{E}[X^{3}]&=1+7g+6g^{2}+g^{3},\\
\mathbb{E}[X^{4}]&=1+15g+25g^{2}+10g^{3}+g^{4}.
\label{momXn}
\end{align}
In addition, the moments-function of $Y$ reads as
\begin{align}
\nonumber
\mathbb{E}[Y^{n}]&\triangleq \mathbb{E}\left[\textstyle \left(\sum^{L}_{i=1}X_{i}\right)^{n}\right]\\
&=\mathbb{E}\left[\sum_{l_{1}+l_{2}+\cdots+l_{L}=n}\frac{n!}{l_{1}!l_{2}!\cdots l_{L}!}\prod^{L}_{i=1}X^{l_{i}}_{i}\right].
\label{momY}
\end{align} 
Expanding the latter general expression for the first four moments and after some tedious yet straightforward manipulations, we arrive at the conditional moments of $Y$, namely
\begin{align}
\nonumber
\mathbb{E}[Y|k]&=\mathbb{E}\left[\left.\textstyle \sum^{k}_{i=1}X_{i}\right\vert k\right]=k \mathbb{E}[X],\\
\nonumber
\mathbb{E}[Y^{2}|k]&=\mathbb{E}\left[\left.\textstyle \left(\sum^{k}_{i=1}X_{i}\right)^{2}\right\vert k\right]=k \mathbb{E}[X^{2}]+k(k-1)\mathbb{E}^{2}[X],\\
\nonumber
\mathbb{E}[Y^{3}|k]&=\mathbb{E}\left[\left.\textstyle \left(\sum^{k}_{i=1}X_{i}\right)^{3}\right\vert k\right]\\
\nonumber
&=k \mathbb{E}[X^{3}]+3 k(k-1)\mathbb{E}[X^{2}] \mathbb{E}[X]\\
\nonumber
&+k(k-1)(k-2)\mathbb{E}^{3}[X],\\
\nonumber
\mathbb{E}[Y^{4}|k]&=\mathbb{E}\left[\left.\textstyle \left(\sum^{k}_{i=1}X_{i}\right)^{4}\right\vert k\right]\\
\nonumber
&=k \mathbb{E}[X^{4}]+4 k(k-1)\mathbb{E}[X^{3}] \mathbb{E}[X]\\
\nonumber
&+6 k(k-1)(k-2)\mathbb{E}[X^{2}]\mathbb{E}^{2}[X]+3 k (k-1)\mathbb{E}^{2}[X^{2}]\\
&+k(k-1)(k-2)(k-3)\mathbb{E}^{4}[X].
\label{momYn}
\end{align} 
The corresponding unconditional moments-function of $Y$ in its general form is presented as
\begin{align}
\mathbb{E}[Y^{n}]\triangleq \sum^{\infty}_{k=1}\mathbb{E}[Y^{n}|k] p^{k-1}(1-p),
\label{momYun}
\end{align}  
which reflects on the case when $k-1$ unsuccessful attempts occurred before the $k^{\rm th}$ successful transmission. By closely observing \eqref{momYn} and utilizing \eqref{momYun}, calculations of the following type arise as
\begin{align}
\nonumber
&\sum^{\infty}_{k=1} k p^{k-1}(1-p) =(1-p)^{-1},\\
\nonumber
&\sum^{\infty}_{k=1} k (k-1) p^{k-1}(1-p) =2p(1-p)^{-2},\\
\nonumber
&\sum^{\infty}_{k=1} k (k-1) (k-2) p^{k-1}(1-p) =6 p^{2}(1-p)^{-3},\\
&\sum^{\infty}_{k=1} k (k-1) (k-2) (k-3) p^{k-1}(1-p) = 24 p^{3}(1-p)^{-4}.
\label{cals}
\end{align} 
Finally, putting altogether the derivations of \eqref{momXn}, \eqref{momYn} and \eqref{cals} in \eqref{varianceAoI}, we reach at \eqref{varianceAoIcf}.

\bibliographystyle{IEEEtran}
\bibliography{IEEEabrv,References}

\begin{thebibliography}{10}
\providecommand{\url}[1]{#1}
\csname url@samestyle\endcsname
\providecommand{\newblock}{\relax}
\providecommand{\bibinfo}[2]{#2}
\providecommand{\BIBentrySTDinterwordspacing}{\spaceskip=0pt\relax}
\providecommand{\BIBentryALTinterwordstretchfactor}{4}
\providecommand{\BIBentryALTinterwordspacing}{\spaceskip=\fontdimen2\font plus
\BIBentryALTinterwordstretchfactor\fontdimen3\font minus
  \fontdimen4\font\relax}
\providecommand{\BIBforeignlanguage}[2]{{%
\expandafter\ifx\csname l@#1\endcsname\relax
\typeout{** WARNING: IEEEtran.bst: No hyphenation pattern has been}%
\typeout{** loaded for the language `#1'. Using the pattern for}%
\typeout{** the default language instead.}%
\else
\language=\csname l@#1\endcsname
\fi
#2}}
\providecommand{\BIBdecl}{\relax}
\BIBdecl

\bibitem{j:RoyYates2021}
R.~D. Yates, Y.~Sun, D.~R. Brown, S.~K. Kaul, E.~Modiano, and S.~Ulukus, ``Age
  of information: {A}n introduction and survey,'' \emph{{IEEE} J. Sel. Areas
  Commun.}, vol.~39, no.~5, pp. 1183--1210, 2021.

\bibitem{j:Yates2020}
R.~D. {Yates}, ``The age of information in networks: {M}oments, distributions,
  and sampling,'' \emph{{IEEE} Trans. Inf. Theory}, vol.~66, no.~9, pp.
  5712--5728, 2020.

\bibitem{j:Kriki2019}
I.~Krikidis, ``Average age of information in wireless powered sensor
  networks,'' \emph{{IEEE} Wireless Commun. Lett.}, vol.~8, no.~2, pp.
  628--631, 2019.

\bibitem{c:Sleem2020}
O.~M. Sleem, S.~Leng, and A.~Yener, ``Age of information minimization in
  wireless powered stochastic energy harvesting networks,'' in \emph{Proc. 54th
  Ann. Conf. Inf. Sci. Syst. (CISS)}, 2020, pp. 1--6.

\bibitem{j:JiaCao2021}
X.~Jia, S.~Cao, and M.~Xie, ``Age of information of dual-sensor information
  update system with {HARQ} chase combining and energy harvesting diversity,''
  \emph{{IEEE} Wireless Commun. Lett.}, vol.~10, no.~9, pp. 2027--2031, 2021.

\bibitem{j:LiuFeng2019}
C.-F. Liu and M.~Bennis, ``Taming the tail of maximal information age in
  wireless industrial networks,'' \emph{{IEEE} Commun. Lett.}, vol.~23, no.~12,
  pp. 2442--2446, 2019.

\bibitem{j:Aziz2020}
M.~K. Abdel-Aziz, S.~Samarakoon, C.-F. Liu, M.~Bennis, and W.~Saad, ``Optimized
  age of information tail for ultra-reliable low-latency communications in
  vehicular networks,'' \emph{{IEEE} Trans. Commun.}, vol.~68, no.~3, pp.
  1911--1924, 2020.

\bibitem{tables}
I.~S. Gradshteyn and I.~M. Ryzhik, \emph{Table of Integrals, Series, and
  Products}.\hskip 1em plus 0.5em minus 0.4em\relax Academic Press, 2007.

\bibitem{j:MiridakisTsif2018}
N.~I. Miridakis, T.~A. Tsiftsis, and G.~C. Alexandropoulos, ``{MIMO} underlay
  cognitive radio: {O}ptimized power allocation, effective number of transmit
  antennas and harvest-transmit tradeoff,'' \emph{IEEE Trans. Green Commun.
  Netw.}, vol.~2, no.~4, pp. 1101--1114, 2018.

\bibitem{b:NIST}
F.~W. Olver, D.~W. Lozier, R.~F. Boisvert, and C.~W. Clark, \emph{NIST Handbook
  of Mathematical Functions}, 1st~ed.\hskip 1em plus 0.5em minus 0.4em\relax
  New York, NY, USA: Cambridge University Press, 2010.

\bibitem{b:Haan2006}
L.~de~Haan and A.~Ferreira, \emph{Extreme Value Theory: An Introduction}.\hskip
  1em plus 0.5em minus 0.4em\relax Springer Series in Operational Research and
  Financial Engineering, 2006.

\bibitem{j:SHIMURA2012}
T.~Shimura, ``Discretization of distributions in the maximum domain of
  attraction,'' \emph{Extremes}, vol.~15, pp. 299--317, 2012.

\bibitem{j:Artzner1999}
P.~Artzner, F.~Delbaen, J.-M. Eber, and D.~Heath, ``Coherent measures of
  risk,'' \emph{Mathematical Finance}, vol.~9, pp. 203--228, 1999.

\bibitem{j:Norton2021}
M.~Norton, V.~Khokhlov, and S.~Uryasev, ``Calculating {CVaR} and {bPOE} for
  common probability distributions with application to portfolio optimization
  and density estimation,'' \emph{Annals of Operations Research}, vol. 299, pp.
  1281--1315, 2021.

\end{thebibliography}

\vfill

\end{document}